\documentclass[a4paper,notoc,hyper]{JHEP}
\usepackage{axodraw}

\title{\boldmath Automatic Integral Reduction for Higher Order Perturbative
Calculations 
}
\author{
Charalampos Anastasiou$^{a}$\footnote{Research supported by the US Department of Energy under contract
DE-AC03-76SF00515},
Achilleas Lazopoulos$^{b}$\footnote{Research supported by the E.U. contract no. HPMD-CT-2001-00105}
\\
$^a$ Theory Group, MS81, 
SLAC, 2575 Sand Hill Rd,
Menlo Park, CA 94025, 
U.S.A.\\
$^b$  Katholieke Universiteit Nijmegen,
Theoretische Fysica,
Postbus 9010,
NL-6500 GL Nijmegen,
The Netherlands
\\
E-mail: \email{babis@slac.stanford.edu},  
\email{lazopoul@sci.kun.nl}
 }
\abstract{ 
We present a program for the reduction of large systems of  
integrals to master integrals. The algorithm was first proposed by Laporta; 
in this paper, we implement it in MAPLE. We also develop two new features 
which keep the size of intermediate expressions relatively small 
throughout the calculation. The program requires modest input information 
from the user and can be used for generic calculations in 
perturbation theory. 
}

\preprint{{SLAC-PUB-10408}}

\begin{document}
\section{Introduction}
\label{sec:intro}

Perturbation theory is an indispensable  calculational  tool in particle 
physics. Methods for perturbative calculations have been developed 
concurrenlty with the introduction of  field theories for 
describing particle interactions. It is not surprising that we
already have very efficient tools which confront experimental data at 
a quantitive level. 

It is evident, however, that the current methods are not suitable 
for computing with sufficient accuracy all the  required cross-sections 
at modern experiments. At the LHC or a future Linear Collider, for example, 
we must study a number of new complicated processes in the Standard Model 
or other theories. In addition, small effects arising at higher orders in 
perturbation theory will become significant in these experiments.  
It is important for such studies to improve or replace methods 
which require substantial human intervention. Ideally, we should develop 
automated  methods applicable to every process, theory, and order in the 
perturbative expansion.  

Two types of computations are generally required for the evaluation of 
cross-sections and decay rates: loop integrations over the momenta of 
virtual particles, and phase-space integrations over the momenta of  
particles in the final state. At higher orders in perturbation theory 
both tasks are hard; this is primarily due to the large number 
of integrals which typically appear. Unfortunately, methods for the 
analytic computation of  loop and phase-space integrals are complicated; 
it is usually unrealistic to attempt a brute-force computation for all terms 
in the matrix-elements. A solution to this problem is   to 
construct algorithms which reduce the number of integrals to a few master 
integrals, and calculate  directly the master integrals only.     

The method of integration by parts (IBP)
for the reduction of loop integrals was introduced in ~\cite{tkachov,ibp}.
Integrals which have common propagators (or, equivalently, belong to the same 
topology) satisfy linear algebraic identities. These identities  can be 
derived with  the IBP method and  can be cleverly combined to 
produce reduction identities to master integrals. 
Gehrmann and Remiddi introduced a new  
class of identities for scalar loop integrals due to their invariance under 
Lorentz transformations~\cite{lorentz}. Lorentz invariance (LI) identities are particularly 
useful for multiloop integrals with many external legs and massive 
propagators.  
Recently, the method of IBP and LI identities 
was extended to phase-space integrals that appear in the evaluation of total 
cross-sections~\cite{htotal,atotal} and various differential 
distributions~\cite{hrap,gammastar,ewk}.

Because of  its conceptual simplicity, 
the IBP method was used to  construct reduction algorithms  
for many classes of multi-loop integrals (see for example~\cite{mincer,betaf,matad,doublebox,crossbox}).  Nevertheless, the construction of such programs  
was 
laborious and a systematic approach to produce reduction identities 
for arbitrary topologies was not available; this was the main reason 
for the rather slow pace of multiloop calculations. 

This situation is now improved due to Laporta, 
who has  proposed  a fully  automated method for the reduction of 
generic loop amplitudes~\cite{laporta}. In contrast to 
earlier approaches, his method does not attempt to derive  reduction 
identities applicaple to all the integrals of  a topology. 
Instead, the aim is to reduce one-by-one the integrals 
by solving a large system of  IBP/LI equations. This is achieved using  
Gauss elimination, after the IBP/LI system is ordered  according to the 
complexity of the equations. 
Starting from the simplest one,  each IBP/LI equation  of the system 
is rearranged following a few algorithmic rules: the  terms of the equation
are assigned a relative weight for their complexity, and the most complicated 
term is then isolated on the left hand side. 
A recursive application of this procedure leads to expressions for 
complicated loop integrals in terms of master integrals. 

The algorithm proposed by Laporta has already been used in a variety of 
calculations (for example in~\cite{koukout,ppqgamma,fourloopbubble,htotal,atotal,lattice1,lattice2,mastrolia,gammastar,ewk,split1,split2,heavylight}). 
However, we have found  that its efficient  implementation in a 
computer program is not trivial. The main difficulties arise from the 
fact that typical multiloop calculations, such as the ones mentioned earlier, 
require an enormous number  of IBP/LI equations ($10^5-10^6$). In the 
process of Gauss elimination the algorithm can produce very large 
expressions; one must optimize for their efficient manipulation. 
    
In this paper, we provide a MAPLE 9~\cite{MAPLE} computer program (AIR)
based on the method of~\cite{laporta}, for the {\bf Automatic Integral Reduction} at higher orders in perturbation theory. 
The user should supply template IBP/LI equations for the integrals of a 
topology, optional information on  the vanishing integrals of 
the topology and the master integrals (if known), and a small number of 
parameters controlling the treatment of large expressions. There is no 
need for advanced knowledge of the  MAPLE platform.  
The input can be supplied with easy to modify text files, and AIR 
can be controled with very simple scripts. 

We believe that theorists
who  do not wish to invest in studying and implementing reduction methods, 
but need to study higher order effects in perturbation theory for 
various physical processes, will find this to be  a valuable tool. 
We also hope that this publication will initiate some activity and exchange 
of ideas on practical issues concerning the implementation of reduction 
algorithms. In this program, we have implemented computational tricks 
wich were developed during practical computations; we hope our program  will 
be improved from the  experience of other users. 

The cost in computer resources grows rapidly with the complexity of the study 
process. It is
inevitable that AIR will fail to solve arbitrarily large 
systems of equations with large number of symbolic parameters 
(corresponding to kinematic scales, dimension, etc). 
However, we do expect AIR to be used 
for many applications in particle phenomenology beyond the current 
state-of-the-art. For this purpose, we have included routines which 
minimize the number of computations during the reductions, mainly by keeping 
the number and the size of the actively processed expressions 
for Gauss-elimination to a minimum.

In Section~\ref{sec:reduction} we review the main features of the algorithm 
of Laporta using the massless one-loop box integral as a pedagogical example. 
In Section~\ref{sec:armi}, we explain the main features of AIR. 
In the rest of the paper we demonstrate the usage of AIR through examples. 
In Section~\ref{sec:nomasking} we reduce the massless one-loop box 
topology with no special algorithms activated for handling large expressions. 
In Section~\ref{sec:mastermasking} we repeat the reduction by activating 
a ``masking''  algorithm which reduces the amount of computations during 
Gauss elimination by storing away the  expressions which get reduced in terms 
of master integrals. In Section~\ref{sec:largemasking} we apply a different
masking algorithm for very large integral coefficients.  
In Section~\ref{sec:pentagons} we show how to reduce topologies with 
a very large number of kinematic scales, by switching off simplification routines. We use the reduction of the massless one-loop pentagon
topology as an explicit example. Finally, we present our conclusions in 
Section~\ref{sec:conclusions}. 

\section{The reduction algorithm}
\label{sec:reduction}
   
In this Section we present the reduction algorithm which is used in 
our program. A detailed description of the algorithm can also be found in 
Ref.~\cite{laporta}. Here we will present its main features using the 
massless one-loop box topology as an explicit example. 
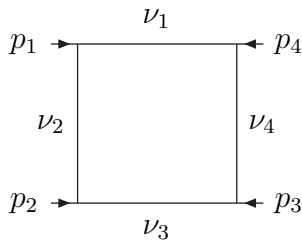
\begin{figure}[h]
\label{fig:box}
\begin{center}
\begin{picture}(110, 110)(0, 0)
\Text(0,80)[]{$p_1$}
\Text(0,20)[]{$p_2$}
\Text(100,80)[]{$p_4$}
\Text(100,20)[]{$p_3$}

\Text(50,10)[]{$\nu_3$}
\Text(50,90)[]{$\nu_1$}
\Text(10,50)[]{$\nu_2$}
\Text(90,50)[]{$\nu_4$}

\Line(20, 20)(80, 20)
\Line(20, 80)(80, 80)
\Line(20, 20)(20, 80)
\Line(80, 20)(80, 80)
\ArrowLine(10, 20)(20, 20)
\ArrowLine(10, 80)(20, 80)
\ArrowLine(90, 80)(80, 80)
\ArrowLine(90, 20)(80, 20)

\end{picture}
\end{center}
\end{figure}
\\
We consider the 
class of integrals: 
\begin{eqnarray}
\label{eq:boxtopology}
B(\nu_1, \nu_2, \nu_3, \nu_4) &=& \int d^dk
\frac{1}{
\left[ k^2\right]^{\nu_1}
\left[ (k+p_1)^2\right]^{\nu_2}
\left[ (k+p_{12})^2\right]^{\nu_3}
\left[ (k+p_{123})^2\right]^{\nu_4}
},
\end{eqnarray} 
where we have introduced the shorthand notation 
$p_{ij \ldots k} = p_i+p_j+ \ldots + p_k$. The terms in the denominator
are raised into positive or negative integer powers $\nu_i$. Zero powers correspond to scalar triangle and bubble 
integrals; negative powers correspond to triangle and bubble integrals  with 
irreducible numerators. The external momenta are all taken to be light-like, $p_1^2=p_2^2=p_3^2=p_{123}^2=0$.
These integrals arise in  one-loop QCD amplitudes for $2 \to 2$ scattering processes (e.g. $gg \to gg$~\cite{sexton}). 

It will be useful to know the values  of the parameters $\nu_i$ for 
which the corresponding integrals vanish (tadpoles, scale-less bubbles).
This information is  not formally required; by solving the IBP equations one will eventually 
find that tadpoles, etc, are indeed vanishing. However, it is more efficient for the reduction 
to utilize the fact that many terms in the IBP equations are zero. 
We find the following vanishing integrals: 
\begin{equation}
B(\nu_1, \nu_2, \nu_3, \nu_4) =0, 
\end{equation}
if 
\begin{equation}
\label{eq:Fzero}
\Theta(\nu_1)+\Theta(\nu_2)+\Theta(\nu_3)+\Theta(\nu_4) < 2,
\end{equation}
or 
\begin{equation}
\Theta(\nu_1)+\Theta(\nu_2) =0,
\end{equation}
or 
\begin{equation}
\Theta(\nu_2)+\Theta(\nu_3) =0,
\end{equation}
or 
\begin{equation}
\Theta(\nu_3)+\Theta(\nu_4) =0,
\end{equation}
or 
\begin{equation}
\label{eq:Lzero}
\Theta(\nu_4)+\Theta(\nu_1) =0,
\end{equation}
where we define $\Theta(x) =1 $ for $x>0$ and $\Theta(x) =0$ for $x \leq 0$.

Now we proceed to  find algebraic equations for the integrals of the box topology. 
An easy way to derive such identities is the IBP method~\cite{ibp,tkachov}; we multiply the integrand with a loop 
or external  momentum and differentiate it with the loop momentum. These total derivatives integrate to zero: 
\begin{equation}
0=\int d^dk \frac{\partial}{\partial k_\mu}
\frac{\eta^\mu}{
\left[ k^2\right]^{\nu_1}
\left[ (k+p_1)^2\right]^{\nu_2}
\left[ (k+p_{12})^2\right]^{\nu_3}
\left[ (k+p_{123})^2\right]^{\nu_4}
},
\end{equation} 
where $\eta=k, k+p_1, k+p_{12}, k+p_{123}$.
We obtain four IBP identities:
\begin{eqnarray}
\label{eq:ibp1}
T_1: 0&=& \left[ s \nu_1 {\bf 1}^+ + (d-\nu_{12334}) - \left( \nu_1 {\bf 1}^+ + \nu_2  {\bf 2}^+ + \nu_4 {\bf 4}^+ 
\right) {\bf 3}^- \right] B \\
\label{eq:ibp2}
T_2: 0&=& \left[ t \nu_2 {\bf 2}^+ + (d-\nu_{12344})- \left( \nu_1 {\bf 1}^+ + \nu_2 {\bf 2}^+ + \nu_3 {\bf 3}^+ 
\right) {\bf 4}^- \right] B \\
\label{eq:ibp3}
T_3: 0&=& \left[ s \nu_3 {\bf 3}^+ + (d-\nu_{11234})- \left( \nu_2 {\bf 2}^+ + \nu_3 {\bf 3}^+ + \nu_4 {\bf 4}^+ 
\right) {\bf 1}^- \right] B \\
\label{eq:ibp4}
T_4: 0&=& \left[ t \nu_4 {\bf 4}^+ + (d-\nu_{12234})- \left( \nu_2 {\bf 1}^+ + \nu_3 {\bf 3}^+ + \nu_4 {\bf 4}^+ 
\right) {\bf 2}^- \right] B 
\end{eqnarray}
where  the action of ${\bf i}^+$ (${\bf i}^-$) increases (decreases)  $\nu_i$ by one in the integral $B$, e.g. 
\begin{equation}
{\bf 3}^{\pm} B = B(\nu_1, \nu_2, \nu_3 \pm 1, \nu_4). 
\end{equation}
Products of operators have a straightforward interpretation, 
e.g. 
\begin{equation}
{\bf 3}^+ {\bf 1}^- B = B(\nu_1-1, \nu_2, \nu_3+1, \nu_4).
\end{equation}
We have also used the shorthand: $\nu_{ijjk \ldots } = \nu_i + 2 \nu_j +\nu_k +\ldots $, and we define the usual Mandelstam variables $s=p_{12}^2, t=p_{23}^2$. 

The IBP Eqs.~\ref{eq:ibp1}-\ref{eq:ibp4} and, optionally,  the results of Eqs.~\ref{eq:Fzero}-\ref{eq:Lzero} 
are sufficient to reduce any integral of the box topology to master integrals by using the algorithm 
of Laporta~\cite{laporta}. In Ref.~\cite{laporta}, the reader can find a detailed and complete description 
of the algorithm; here we intend to emphasize its salient features. The user is not required to have knowledge 
of the algorithm, however, some  familiarity will be beneficial. 
We will describe the algorithm by tracing the first steps of our code when solving the box topology.   
For concretness, we will stop when the integral B(1, -1, 1, 0) is reduced in terms of master integrals. 

\begin{itemize}
\item {\it Seed generation}: 
The program starts with  the simplest list  $(\nu_1, \nu_2, \nu_3, \nu_4)$ for which $B(\nu_1, \nu_2, \nu_3, \nu_4)$ is 
not vanishing,
\begin{equation}
S_1 : (\nu_1, \nu_2, \nu_3, \nu_4) = (1, 0, 1, 0).
\end{equation} 
$S_1$ will be our first `seed' for generating identities from the topology IBP equations $T_1,\ldots, T_4$ 
(Eqs.~\ref{eq:ibp1}-\ref{eq:ibp4}), which we use as templates. 

\item {\it Identities generated from the template IBP equations}:
Our first identity is Eq. $T_1$ substituting the  values of $\{\nu_i \}$ found in $S_1$: 
\begin{equation}
E_1^1: sB(2, 0, 1, 0) + (d-3) B(1, 0, 1, 0)=0.
\end{equation} 
In $E_1^1$, we have already used our knowledge  for the vanishing integrals of the 
topology (Eqs.~\ref{eq:Fzero}-\ref{eq:Lzero}). 

The above equation can be recast  to express one of the two integrals in terms of the other.
We would like to use such equations to express more complicated integrals in terms of simpler ones and, 
finally, in terms of the master integrals.  It is therefore necessary to introduce criteria for the complexity of the 
integrals; the most complicated should receive first priority and will be isolated in the left hand side.

\item {\it Integral priority criteria}: We check on three parameters in order to isolate the most complicated integral. 
First we select the integrals with the largest number of propagators:
\begin{equation}
{\cal N}_{prop} =  \sum_i \Theta(\nu_i).  
\end{equation} 
If more that one integral has the maximum ${\cal N}_{prop}$, we select the one
with the largest sum of positive indices $\nu_i$
\begin{equation}
{\cal N}_+ = \sum_i \Theta(\nu_i) (\nu_i-1).
\end{equation}  
If more than one integral has the maximum values of $N_{prop}$ and $N_{+}$, 
we select the one with the largest sum for the magnitutes of negative 
indices $\nu_i$ 
\begin{equation}
{\cal N}_- = -\sum_i \Theta(-\nu_i) \nu_i.
\end{equation}
If still there is an ambiguity, we randomly choose one of the integrals which has survived all 
three criteria. 

\item {\it Rearranging the identities}: 
Following the previous criteria, we find that B(2, 0, 1, 0)  is the most
complicated integral in $E_1^1$. We then rearrange the identity to produce: 
\begin{equation}
E_1^1: B(2, 0, 1, 0)= \frac{3-d}{s} B(1, 0, 1, 0).
\end{equation} 
We proceed, in the same manner, with the remaining identities $T_2, T_3, T_4$ for $S_1$. We obtain:
\begin{equation}
E_1^2: B(2,0,1,-1) = (d-2) B(1,0,1,0) - B(1,0,2,-1), 
\end{equation}
\begin{equation}
E_1^3: B(1, 0, 2, 0)= \frac{3-d}{s} B(1, 0, 1, 0),
\end{equation} 
and 
\begin{equation}
\label{eq:subi}
E_1^4: B(2,-1,1,0) = (d-2) B(1,0,1,0) - B(1,-1,2,0). 
\end{equation}

\item {\it Seed priority criteria:} We have now processed all IBP equations for the first 
seed $S_1$. It is therefore necessary to choose a new seed to obtain more identities. It is important to choose seeds that are most likely to produce 
equations coupled with the ones which have been 
processed earlier. For this purpose, we could select the seeds with the opposite priorities than the 
integral priorities, i.e. the seed with succecively minimum values for 
$[{\cal N}_{prop}, {\cal N}_-, {\cal N}_+]$. However, 
the rules for choosing the seeds are mostly empirical and require 
some experimentation. In fact, the order for applying  the criteria 
for minimum ${\cal N}_-$ and ${\cal N}_+$ can be judiciously chosen 
according to the class of integrals that the user needs to compute. 
For example, we could now pick either $(1, -1, 1, 0)$ or 
$(2, 0, 1, 0)$ as the next seed. Since our goal is to compute 
$B(1, -1, 1, 0)$ it is better to choose:
\begin{equation}
S_2 : (\nu_1, \nu_2, \nu_3, \nu_4) = (1, -1, 1, 0).
\end{equation} 
which in the IBP equations generates integrals with the same structure 
as the one we want to solve. Our program generates the seeds automatically;
the user must provide the range of  ${\cal N}_{prop}, 
{\cal N}_-, {\cal N}_+$ as input. It is relatively straightforward to decide 
the values for these parameters by inspecting the integrals that are required
in the study process.

\item {\it Substitutions and Gauss-Elimination:} We now find a new feature in Eq. $T_1$ for the seed $S_2$: 
\begin{equation}
E_2^1: sB(2, -1, 1, 0) + (d-2) B(1, -1, 1, 0) =0. 
\end{equation}  
The integral $B(2, -1, 1, 0)$ is  isolated at the left hand side (lhs) of 
a previous equation ($E_1^4$). In such cases, we eliminate 
the known integral from the equation. Substituting Eq.~\ref{eq:subi}, and applying the integral priority criteria, 
we have: 
\begin{equation}
\label{eq:bsubi}
E_2^1: B(1, -1, 2, 0) = \frac{d-2}{s} B(1,-1,1,0) + (d-2) B(1, 0, 1, 0)
\end{equation}

\item {\it Back-substitution:} $E_2^1$ is solved in terms of an integral that can be substituded back to
$E_1^4$. We can now see how the Laporta algorithm works in practice; 
by adding new equations to the already solved 
equations we form new sub-systems of coupled equations wich eliminate previously unknown integrals. In our example, performing the  substitution of $E_2^1$  
into $E_1^4$ we obtain, 
\begin{equation}
E_1^4: B(2, -1, 1, 0) = \frac{2-d}{s} B(1, -1, 1, 0).
\end{equation}
We process two more equations for the seed $S_2$: 
\begin{eqnarray}
E_2^2: && B(2, -1, 1, -1) = -tB(1, 0, 1, 0) 
+(d-1) B(1, -1, 1, 0) \nonumber \\ 
&& \quad + B(1, 0, 1, -1) - B(1, -1, 2, -1),
\end{eqnarray}
and 
\begin{equation}
\label{eq:target}
E_2^3: B(1, -1, 1, 0) = -\frac{s}{2} B(1, 0, 1, 0). 
\end{equation}    
In the last equation we have computed the integral that we wanted in terms 
of a simpler one: $B(1, 0, 1, 0)$. It is clear from the previous equations 
that B(1, 0, 1, 0) is a master integral.
\end{itemize}

In summary, the algorithm requires the succecive generation of identities  
with terms of increasing complexity. The newly added  equations 
usually contain terms which are also found 
in equations generated at earlier stages; this produces small subsystems of 
coupled algebraic identities. A series of substitutions  
diagonilizes these algebraic subsystems and yields complicated integrals 
expressed in terms of master integrals.  The algorithm is a clever 
implementation of Gauss elimination. It exploits 
the fact that Feynman integrals can be ordered according to 
very simple criteria.  

We  demonstrated how the algorithm reduces a number of integrals belonging 
to the massless one-loop box topology. 
However, there was no step in the previous reduction that depended 
on the specifics of the topology.  Therefore, this algorithm is suitable 
for the reduction of generic multiloop integrals or, more generally, of 
parametric functions which satisfy coupled algebraic 
identities (e.g. hypergeometric functions).

\section{Features of AIR}
\label{sec:armi}

In this Section we describe the basic functions of our program. The program 
is included as a gzipped and tarred file in the source submission of
the electronic preprint for this paper, and  can also be downloaded from  
Ref.~\cite{airlocation}. It is convenient to unzip and 
untar the distribution file in a directory where AIR can be located 
permanently. 
\begin{verbatim}
/home> tar -zxvf air.tar.gz 
AIR/
AIR/main.map
AIR/BOXA/
AIR/BOXA/input_boxa.map
AIR/BOXA/script_boxa.map
AIR/BOXB/
AIR/BOXB/input_boxb.map
AIR/BOXB/script_boxb.map
AIR/BOXC/
AIR/BOXC/input_boxc.map
AIR/BOXC/script_boxc.map
AIR/Pentagon/
AIR/Pentagon/input_pentagon.map
AIR/Pentagon/script_pentagon.map
AIR/Pentagon5/
AIR/Pentagon5/input_pentagon.map
AIR/Pentagon5/script_pentagon.map
\end{verbatim} 

The distribution includes the program file {\it main.map},  
input files {\it input\_$\cdots$.map}, as well as MAPLE scripts {\it script\_$\cdots$.map} for the example reductions in the rest 
of the paper. The program  consists of MAPLE routines for generating seeds for the 
template IBP/LI identities, finding integral priorities, generating the 
IBP equations from the seeds, performing Gauss-elimination, masking large integral coefficients and reduced expressions, performing nested substitutions,
and collecting the results. The function of the more important routines 
will be detailed in the following Sections.

In typical multiloop computations, a 
large number of identites should be processed. 
A database system is therefore required 
to access, modify, and store the equations.  
We have implemented a rather simple database system, where 
each IBP equation is stored in a single file; the name of the file is 
determined by the integral on the lhs of the equation. 
We also create separate auxiliary files which serve to  
point to the equations in the IBP system where a particular integral can be 
found. Our database system is very robust; however, it creates a rather 
extended tree of directories wich  usually contain  very short ASCII files.

The program can perform  very complicated multiloop reductions.  
It is often possible to simplify all the terms in  the IBP equations
as they get substituted and rearranged for Gauss elimination.  
However, if the topology depends on many  kinematic scales, or 
the IBP equations are loosely coupled (creating large subsystems of equations 
before they get diagonalized), or the values for ${\cal N}_{prop}, 
{\cal N}_{\pm}$ are large, it may not be feasible to perform all 
simplifications within acceptable times or the available  memory.  
We have implemented  two algorithms to perform the reductions 
efficiently and  reduce the amount of computations; the algorithms  
can be used independently or in conjuction. 

The first algorithm  masks subexpressions which are  
reduced in terms of master integrals. The program detects the reduced 
expressions, stores them in files, and replaces them by an 
indexed symbol. Thus, the masked expressions are protected from subsequent
 manipulations during Gauss elimination. This feature is implemented 
recursively; whenever a new expression is written 
in terms of masked expressions and/or master integrals,  the 
new expression is also masked. At the end of the reduction, a series of 
nested substitutions is required in order to rewrite 
the masked expressions in terms of the master integrals. We will discuss the 
required nested substitutions later; for now, we should note that the 
masking algorithm reduces significantly the amount of computations  during 
the process of Gauss elimination. 
The algorithm requires that the master integrals of the topology are 
known. To determine them, one can perform a less involved reduction 
for relatively small values of  ${\cal N}_{\pm}$ without using the masking 
algorithm. When the master integrals are found, the user can repeat the 
reduction for larger values of ${\cal N}_{\pm}$, activating the masking 
algorithm. 

The second algorithm aims to reduce the size of the equations by 
masking all integral coefficients which are lengthier than a user-defined  
maximum value.  During Gauss-elimination, however, some integral 
coefficients vanish;  the elimination cannot take place if the coefficients 
contain masked expressions.  To solve this problem we check numerically for 
cancelations. The user needs to provide as input, 
numerical values for all the parameters (kinematic scales, dimension) 
which enter in the symbolic expressions for the integral coefficients
of the IBP/LI equations. The masking algorithm  substitutes 
these numerical values and stores both the numerical result and the 
symbolic expression for the coefficients. The program determines  if a 
coefficient is zero by inspecting the numerical result, thus, avoiding 
complicated symbolic manipulations.   
To ensure that  cancelations are not accidental, the program 
can perform  the numerical testing of the values of the coefficients  
for more than one choices of numerical values for the kinematic parameters 
and the dimension. The analytical value of the lengthy 
coefficients is computed at the end of the reduction, and only for the 
integrals that are required for practical purposes. 

The purpose of the two algortihms is to remove 
complications from the symbolic manipulation of very 
large expressions. However, after Gauss-elimination we must still perform 
computations which were defered by using the masking algorithms, i.e.  
we must perform a series of nested symbolic substitutions for the used alias 
symbols in order to compute the masked expressions explicitly in terms of 
the kinematic parameters, the dimension, and the master integrals. 
It is possible to 
imagine that this additional computation is as difficult as using the program 
without the masking algorithms, where all substitutions take 
place explicitly  during Gauss elimination. 
This is not the case; usually, only a 
fraction of the total number of masked expressions is required 
for the integrals that appear in the matrix-elements of a physical process. 
For example, the integral $B(2, -1, 1, 0)$ of the previous Section does 
not appear in the computation of e.g. $gg \to gg$, however, it appears 
in the IBP equations. By using the masking algorithms, we avoid
computing the masked expressions for many such 
integrals. 

The remaining nested substitutions can still be challenging  for 
very complicated problems. One can resort to  tricks such as 
expanding in the dimension parameter~\cite{mincer} or in 
kinematic parameters (e.g. electron mass in Bhabba scattering), if 
this is justified from the physics of the process. However, this is 
rarely needed; there are many processes where we can perform the 
substitutions without giving up on a valid evaluation of the integral 
coefficients for all values of the kinematic parameters and the dimension. 
AIR includes general purpose routines for a straightforward computation of 
recursive substitutions; these routines attempt a brute-force symbolic 
simplification of  all the intermediate expressions.
It also provides the option to switch off simplification 
of expressions that exceed a maximum length or, if necessary,  
to transfer the most complicated substitutions to another platform, 
e.g. FORM~\cite{form}. 

In the following Sections we perform four example reductions which 
can serve as a tutorial for using AIR and its main features. A technical 
description of the AIR routines can be found in~\cite{airlocation}. 
The source code of the program is openly distributed; the users are free 
to modify it. The authors will be greatful to receive suggestions and 
constructive feedback. 
 
\section{Reduction with no masking}
\label{sec:nomasking}

We now perform the reduction of the massless one-loop box topology. 
In this Section we do not activate any of the two  masking algorithms.
The input file and the corresponding script for the reduction can be 
found in the directory:
\begin{verbatim} 
/home/AIR/BOXA
\end{verbatim}  
The input file for the reduction is named {\tt input\_boxa.map}. 
It contains variables which are used globally by AIR. These are:
\begin{verbatim}
ibp_equations:= [
-nu3*B(nu1-1,nu2,nu3+1,nu4)-nu4*B(nu1-1,nu2,nu3,nu4+1)
-nu2*B(nu1-1,nu2+1,nu3,nu4)+nu3*s*B(nu1,nu2,nu3+1,nu4)
+(-nu3-nu2-2*nu1+d-nu4)*B(nu1,nu2,nu3,nu4), 
nu4*t*B(nu1,nu2,nu3,nu4+1)-nu3*B(nu1,nu2-1,nu3+1,nu4)
+(d-2*nu2-nu3-nu1-nu4)*B(nu1,nu2,nu3,nu4)
-nu4*B(nu1,nu2-1,nu3,nu4+1)-nu1*B(nu1+1,nu2-1,nu3,nu4), 
-nu1*B(nu1+1,nu2,nu3-1,nu4)+(-nu2+d-nu4-nu1-2*nu3)
*B(nu1,nu2,nu3,nu4)-nu4*B(nu1,nu2,nu3-1,nu4+1)
-nu2*B(nu1,nu2+1,nu3-1,nu4)+nu1*s*B(nu1+1,nu2,nu3,nu4), 
-nu1*B(nu1+1,nu2,nu3,nu4-1)-nu2*B(nu1,nu2+1,nu3,nu4-1)
-nu3*B(nu1,nu2,nu3+1,nu4-1)+nu2*t*B(nu1,nu2+1,nu3,nu4)
+(-nu3+d-nu2-nu1-2*nu4)*B(nu1,nu2,nu3,nu4)
]:
\end{verbatim}
\begin{itemize}
\item This is a list of template IBP identities (Eqs.~\ref{eq:ibp1}-
~\ref{eq:ibp4}) for the box topology. The program reads off some 
additional implicit definitions from the structure of the IBP equations. 
For example, it is now defined that the name of the topology is 
{\tt ``B''} and the powers of the propagators are defined through the 
variables {\tt nu1, nu2, nu3, nu4}. 
\end{itemize}
\begin{verbatim}
ZERO_TOPOLOGIES:=[
ThetaF(nu1) + ThetaF(nu2) + ThetaF(nu3) +ThetaF(nu4) < 2,
ThetaF(nu1) +ThetaF(nu2) =0,
ThetaF(nu2) +ThetaF(nu3) =0,
ThetaF(nu3) +ThetaF(nu4) =0,
ThetaF(nu4) +ThetaF(nu1) =0,
NULL]:
\end{verbatim}
\begin{itemize}
\item This is a list of statements (Eq.~\ref{eq:Fzero}-\ref{eq:Lzero})
which undergo boolean evaluation when the propagator powers  are 
substituted by integers.  If any of the statements is true, then the 
corresponding integral is set to zero. 
\end{itemize}

\begin{verbatim}
MASTERS:=[]:
\end{verbatim}
\begin{itemize}
\item This variable activates the algorithm for masking reduced expressions
to master integrals. It contains a list of known master integrals. In this 
example, we do not want to activate the algorithm; therefore we define 
the above variable to be an empty list. 
\end{itemize}
\begin{verbatim}
check_values:=[]:
\end{verbatim}
\begin{itemize}
\item  This variable activates the masking algorithm for integral 
coefficients which exceed a maximum value. It should contain numerical 
values for all the kinematic parameters and the dimension in the 
IBP/LI equations. For this example, we do not want to activate the 
masking algorithm and we set the variable to an empty list. 
\end{itemize}
\begin{verbatim}
MAXLENGTH:=1000: 
\end{verbatim}
\begin{itemize} 
\item 
This variable is used by the masking algorithm for large integral 
coefficients. It defines the maximum length for a coefficient in order not 
to get masked. The length of an expression is measured by the number of 
characters in the expression and is determined by a MAPLE routine. The value
of the variable is irrelevant if the {\tt check\_values} list is empty.
\end{itemize}
\begin{verbatim}
MAXSIMPLIFY:=10^10:
\end{verbatim}
\begin{itemize}
\item This variable is used by the routines which perform nested substitutions 
for the masked expressions and the routines which display the final results. 
It sets a maximum length for the expressions that MAPLE is allowed to 
simplify. Larger expressions get substituted but not simplified. 
\end{itemize}
\begin{verbatim}
MAPLEMAXSUB:=10^10:
\end{verbatim}
\begin{itemize}
\item This variable is used by the routines which perform  nested 
substitutions of masked expressions, and the routines wich display 
the results. It sets a maximum length for the expressions that MAPLE is 
allowed to substitute. Larger expressions are not written explicitly and 
are kept masked. In this reduction we want all coefficients to be explicit, 
and we set the value  of the variable to a practically unreachable value.

\end{itemize}
\begin{verbatim}
VERBOSE:=FALSE:  
\end{verbatim} 
\begin{itemize} 
\item This variable is used to display information about the progress of the 
program. 
If set to TRUE, the program outputs  on the screen the seed that 
is processing, or, after Gauss-elimination is comleted, 
the index of the masked expression that is evaluating. 
\end{itemize}

The file {\tt script\_boxa.map}  contains all the calls to  
AIR for reducing the box topology. We can run the script from the 
shell command line: 
\begin{verbatim}
/home/AIR/BOXA> maple script_boxa.map
\end{verbatim}
It will be more instructive for this first application to call the 
routines interactively from within the MAPLE platform. We fisrt 
launch MAPLE, 
\begin{verbatim}
/home/AIR/BOXA> maple
\end{verbatim}
and load the input for the topology and the main program: 
\begin{verbatim}
> currentdir(``/home/AIR/BOXA''):
> read ``input_boxa.map'':
> read ``/home/AIR/main.map'':
\end{verbatim} 
The user should now perform the following tasks:
\begin{itemize}
\item {\it Seed generation}: We create a list with sets of integers for 
deriving IBP equations from the template equations of
{\tt ibp\_equations}. This is accomplished by calling the routine
\\ \\
{\bf 
SEEDGEN(``filename'', maxtop, [minNprop,maxNprop], [minNminus, maxNminus], [minNplus, maxNplus]) }; \\
\\
The first argument is the filename where the seeds will be written.  
The next argument, {\tt maxtop}, is a list with integers  $\{i\}$
denoting the propagators raised to positive powers $\{\nu_i\}$ 
in the seed with the highest priority.  
For example, if we are inerested in reducing all the integrals  
of the box topology and all its subtopologies we should set 
${\tt maxtop} = [1,2,3,4]$, indicating that 
all $\nu_1,\nu_2,\nu_3,\nu_4$ can appear with positive values. 
If we only required integrals of  e.g. a  $t$-channel triangle subtopology, we could 
set ${\tt maxtop} = [1,3,4]$, indicating that seeds with positive 
$\nu_2$ do not need to be included in the reduction. 
The next three entries determine the range of 
${\cal N}_{prop}$, ${\cal N}_-$, and ${\cal N}_+$ repsectively. 
These values are mostly empirical. A rule of thumb is that one should 
generate seeds wich include the indices of the most comlicated integrals 
to be reduced, as well as the complete tower of integrals 
with lower priorities. To give a concrete example, we will generate seeds 
for the integrals that appear in the one-loop $gg \to gg$ amplitude. 
We find integrals from  all subtopologies (bubble and triangles) 
of the box topology, therefore we  set $2 \leq {\cal N}_{prop} \leq 4$. 
We also find that all integrals have ${\cal N}_+=0$ (there are no squared 
propagators in the amplitude), and can have up to $4$ powers of irreducible 
numerators: $0 \leq {\cal N}_{-} \leq 4$. To emphasize differences in the 
running times for  various algorithms of the program we will extend the 
latter interval to $0 \leq {\cal N}_{-} \leq 10$. 
\begin{verbatim}
> SEEDGEN(``seeds.map'', [1, 2, 3, 4], [2, 4], [0, 10], [0, 0]);
\end{verbatim}  
The routine produces a list of ordered seeds in the file 
{\tt /home/AIR/BOXA/seeds.map}. 

\item {\it Gauss elimination}: We generate the IBP equations from the 
seeds and perform Gauss elimination by calling  the routine\\ \\
{\bf 
Reducer(``seeds\_file'', ``monitor\_file'', ``RESULTS\_DIR'');
}
\\ 
\\
The first argument is the name of the file with the seeds for the reduction
(as it was produced with {\tt SEEDGEN}). 
The next argument is the name of a file  which the program updates 
with the processed seeds; it serves to monitor the progress of the program. 
The last argument provides a directory path where the program can deposit 
the database with the IBP equations. For our example, we 
type 
\begin{verbatim}
> Reducer(``seeds.map'', ``calc.map'', ``.'');
\end{verbatim}
It is useful to inspect the {\tt /home/AIR/BOXA} directory. 
The program has created a number of subdirectories ({\tt B\_1\_3, 
B\_1\_3\_4}, etc) which correspond to the non-vanishing 
subtopologies of the box topology. In these subdirectories, AIR stores 
the IBP equations after they have been rearranged to isolate the most 
complicated integral at the lhs of the equation. For example, 
Eq.~\ref{eq:target} is stored in 
\begin{verbatim}
/home/AIR/BOXA/B_1_3/B_1_-1_1_0.map
\end{verbatim}
The first part of the file path {\tt /home/AIR/BOXA/} corresponds to 
the directory in the third argument of the {\tt Reducer} command. 
The second part of the file path {\tt B\_1\_3} is created from 
the integers $\{i\}$ for which the powers $\{ \nu_i \}$ are 
positive. The ending of the file path is created from the indices $\{\nu_i \}$
of the integral. 
 
\item {\it Collecting the results}: Essentially the reduction 
is now complete; the reduced integrals can be found in the files of the 
database tree for the IBP equations. 
Inspecting some of the equations we find that many integrals are 
reduced in terms of three master integrals: 
B(1, 0, 1, 0), B(0, 1, 0, 1), and B(1, 1, 1, 1). 
However, we also find integrals which are not fully reduced, 
e.g. the integral B(2, 1, 1, -11). 
It is usually observed that integrals 
with the same indices as in the seeds are fully reduced. 
Motivated from this observation, we have written a routine for collecting 
in a separate directory the seed integrals:
 \\ \\
{\bf 
tidy\_list(``seeds\_file'', ``RESULTS\_DIR'' );
}\\ \\
The first argument is a file with seeds (as generated by {\tt SEEDGEN}). 
The last argument is the directory where the program has placed 
the results (as in the third argument for {\tt Reducer}). 
For our example, we type:
\begin{verbatim}
> tidy_list(``seeds.map'', ``.'');
\end{verbatim}
The routine has created a subdirectory, named {\tt RESULTS}, 
 wich contains the expressions for the seed integrals only. 
For example, one can find the integral B(1, -4, 1, 1) in  the file 
{\tt /home/AIR/BOXA/RESULTS/B\_1\_3\_4/B\_1\_-4\_1\_1.map}. 
We should note that when the masking algorithms are activated, 
the {\tt tidy\_list} routine  also performs the required  nested 
substitutions for expressing the coefficients of the 
master integrals in terms of the kinematic parameters 
and the dimension. However, it does so only for coefficients with smaller 
size than the {\tt MAPLEMAXSUB} value. It is possible to achieve a fast 
execution of the {\tt tidy\_list} routine if we  choose  a low value (100-500) 
of {\tt MAPLEMAXSUB}. 
The task of computing large coefficients is postponed further, and is 
performed by a new routine which we shall describe shortly. This routine is 
customized to disentagle the newest masking, which has a simpler structure 
than the masking of Gauss-elimination, very efficiently.

\item {\it Reading the results interactively:}
A useful routine for reading  the reduced integrals from within the 
MAPLE environemnt is 
\\ 
\\
{\bf 
show\_int(integral, RESULTS\_DIR);
}
\\
\\
The first argument is the  required integral, and the second is the directory 
of the reduction (as in {\tt Reducer}).
For example, 
\begin{verbatim}
> show_int(B(1, -4, 1, 1), ``.'');
\end{verbatim}
outputs the expression for the required integral in terms of master 
integrals. 
This routine can be used for purposes of interfacing the results of the 
reduction to other programs that users develop for calculating 
matrix-elements. We should note that if the masking of master integrals is activated
and the value of {\tt MAPLEMAXSUB} is small, {\tt show\_int} will return the 
wanted integral as a linear combination of master integrals but with masked 
coefficients. The routine {\tt show\_full\_int} displays the unmasked result.
\end{itemize}
We have described the basic variables and routines  of AIR by performing 
the reduction of the box topology. It is worth noting that the total 
running time for the three main routines (SEEDGEN, Reducer, tidy\_list) 
is approximately two minutes on a $1.6 {\rm GHz}$ processor.

\section{Masking reduced expressions}
\label{sec:mastermasking}

In this section we repeat the reduction of the one-loop massless box 
topology providing the known master integrals as input to the program. 
The algorithm uses this information to find parts of expressions which are 
reduced to master integrals, and masks them. The reduction proceeds faster, 
having replaced  the masked expressions  by indexed symbols $K(i)$.  
When the step of Gauss elimination is completed, only the indexed symbols 
in the expressions of the seed integrals need to be  computed explicitly. 

The reduction is performed in the directory {\tt /home/AIR/BOXB}.  
The program  allows, in principle,  activation of the masking algorithm  
for master integrals without changing directory. However, we
recommend performing { 
 the reduction in  a new  directory if new values for the 
variables {\tt ZERO\_TOPOLOGIES}, {\tt 
ibp\_equations}, {\tt MASTERS}, 
{\tt check\_values}, and {\tt MAXLENGTH} are required. This will 
prevent  possible integral misidentifications  caused from creating the 
database of IBP equations with inconsistent values for these variables.  
  
The input file {\tt input\_boxb.map} is modified to activate the 
masking algorithm for master integrals. In particular, the variable 
for the master integrals is not empty; we have defined
\begin{verbatim}
MASTERS:=[
	B(1,0,1,0), 
	B(0,1,0,1),
	B(1,1,1,1)]:
\end{verbatim} 
These are the  master integrals that we found in the first 
reduction.

We now execute the  MAPLE script
in {\tt script\_boxb.map}: 
\begin{verbatim}
currentdir("/home/AIR/BOXB"):
read "input_boxb.map":
read "/home/AIR/main.map":
SEEDGEN("seeds.map", [1, 2, 3, 4], [2, 4], [0, 10], [0, 0]):
Reducer("seeds.map", "calc.map", "."):
tidy_list("seeds.map", "."):
\end{verbatim}
The script is executed in less than a  minute; this is approximately 
50\% faster than without the masking algorithm. 
The routine {\tt Reducer} has created   a tree of  subdirectories 
{\tt ICED/ICED\#/KEXPR}, where the masked expressions are stored in files 
named {\tt kexpr\_i.map}. 
The expressions are replaced in the reduction by the symbol $K(i)$, 
where $i$ is an integer index. 

The masked expressions are defined  recursively; 
it is therefore necessary 
to perform a series of  nested substitutions before we obtain their 
explicit form in terms of master integrals. This task is performed by the 
{\tt tidy\_list} routine, which stores explicit results for the masked 
expressions in the directory {\tt /home/AIR/BOXB/KMELT}. The user 
is not required to know the details about the file structure for the masked 
expressions; the integrals in the directory {\tt RESULTS}  are fully 
evaluated in terms of master integrals and can be accessed as before. 
 
\section{Masking large integral coefficients}
\label{sec:largemasking}
  
In this Section we describe  the function of the  algorithm which 
masks  large integral coefficients in the IBP equations. 
We perform the reduction
in the directory {\tt /home/AIR/BOXC}, which contains the input file 
{\tt input\_boxc.map} and the appropriate MAPLE script 
 {\tt script\_boxc.map}. 
We have modified the input variables: 
\begin{verbatim}
MASTERS:=[]:
check_values:=[
[s=1,t=-0.12,d=3.3],
[s=-1.2, t=12.2, d=42],
[s=1.82, t=-0.345, d=-28.1]
]:
MAXLENGTH:=50: 
\end{verbatim}
In this run we have deactivated the algorithm for masking 
reduced expressions. Instead, we have provided a list of 
numerical values for the kinematic parameters and the dimension which 
activates  the algorithm for masking the integral coefficients with length 
bigger that the value of the {\tt MAXLENGTH} variable.
The MAPLE commands for the reduction are collected in the file {\tt  script\_boxc.map}, 
\begin{verbatim}
restart;
currentdir("/home/AIR/BOXC"):
read "input_boxc.map":
read "/home/AIR/main.map":
SEEDGEN("seeds.map", [1, 2, 3, 4], [2, 4], [0, 10], [0, 0]):
Reducer("seeds.map", "calc.map", "."):
tidy_list("seeds.map", "."):
\end{verbatim}
The script is executed in about 90 seconds; this is slower than previously. 
In general, masking large coefficients is not as fast as masking the reduced 
expressions to master integrals. However, in reductions 
of complicated IBP systems, such as in mulitloop crossed topologies,
we have found that this algorithm is indispensable. 
In  extremely complicated problems it is required that both masking 
algorithms are activated. 

In the directory {\tt ICED} we find two new subdirectories: 
{\tt EXPR} and {\tt NUM}. The routine {\tt Reducer}
saves the expressions for the masked large integral coefficients in the 
first directory. 
In the second directory, it stores the 
numerical values of the same coefficients for the choices of the parameters in 
 {\tt check\_values}. Expressions in the directory 
{\tt EXPR} are defined recursively through other masked expressions. 
However, in the directory {\tt NUM} they are explicitly evaluated for the 
special input values of the parameters. 
The program sets an expression to zero in the IBP equations if its numerical
values is zero 
for all input  choices in {\tt check\_values}. 
It is important to provide ``sufficiently random'' lists of numerical 
values for the parameters in order to reduce the risk for accidental 
cancelations. It is important to note that 
floating point numbers in {\tt check\_values} are automatically 
{\emph  converted} to fractional numbers when used in the program. Therefore, 
the numerical evaluation of coefficients is exact (using integer arithmetics), 
avoiding complications due to rounding. 

Finally, the results are collected in the directory {\tt RESULTS} using  
the routines {\tt tidy\_list} and  {\tt show\_int}.  We should stress, that 
the results 

\section{Reduction of one-loop pentagons}
\label{sec:pentagons}

In this section we perform the reduction of the one-loop pentagon 
with massless propagators. This is an example of a topology with 
many kinematic parameters. As is common in such topologies, 
the expressions for the reduced  integrals are large. 
We will therefore use this example to demonstrate the options in AIR
for  dealing  with large expressions.
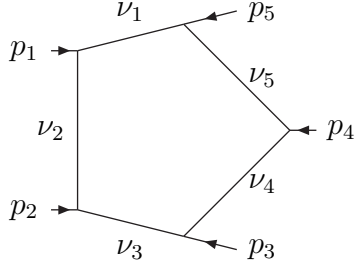
\begin{figure}[h]
\label{fig:pentagon}
\begin{center}
\begin{picture}(110, 110)(0, 0)
\Text(0,80)[]{$p_1$}
\Text(0,20)[]{$p_2$}
\Text(90,95)[]{$p_5$}
\Text(120, 50)[]{$p_4$}
\Text(90,5)[]{$p_3$}

\Text(40,5)[]{$\nu_3$}
\Text(40,95)[]{$\nu_1$}
\Text(10,50)[]{$\nu_2$}
\Text(90,30)[]{$\nu_4$}
\Text(90,70)[]{$\nu_5$}

\Line(20, 20)(20, 80)
\Line(20, 20)(60, 10)
\Line(20, 80)(60, 90)
\Line(60, 10)(100, 50)
\Line(60, 90)(100, 50)

\ArrowLine(10, 20)(20, 20)
\ArrowLine(10, 80)(20, 80)
\ArrowLine(110, 50)(100, 50)
\ArrowLine(80, 5)(60, 10)
\ArrowLine(80, 95)(60, 90)

\end{picture}
\end{center}
\end{figure}
\\
We consider the 
class of integrals: 
\begin{eqnarray}
\label{eq:Vtopology}
V(\nu_1, \nu_2, \nu_3, \nu_4, \nu_5) &=& 
\nonumber \\
&& \hspace{-3cm}\int d^dk
\frac{1}{
\left[ k^2\right]^{\nu_1}
\left[ (k+p_1)^2\right]^{\nu_2}
\left[ (k+p_{12})^2\right]^{\nu_3}
\left[ (k+p_{123})^2\right]^{\nu_4}
\left[ (k+p_{1234})^2\right]^{\nu_5}
},
\end{eqnarray} 
where $p_{12345}=0$ and $p_1^2=p_2^2=p_3^2=p_4^2=p_5^2=0$. 
The IBP equations for the pentagon topology are:
\begin{eqnarray}
&& \hspace{-1cm}
\nu_1 {\bf 1^+} s_{12} + \nu_5 {\bf 5^+} s_{34} + (d-\nu_{123345}) 
-\left( \nu_1 {\bf 1^+} + \nu_2 {\bf 2^+} + \nu_4 {\bf 4^+} + \nu_5 {\bf 5^+}
 \right) {\bf 3^-} = 0  \\
&& \hspace{-1cm}
\nu_2 {\bf 2^+} s_{23} + \nu_1 {\bf 1^+} s_{45} + (d-\nu_{123445}) 
-\left( \nu_1 {\bf 1^+} + \nu_2 {\bf 2^+} + \nu_3 {\bf 3^+} + \nu_5 {\bf 5^+}
 \right) {\bf 4^-} = 0  \\
&& \hspace{-1cm}
\nu_3 {\bf 3^+} s_{34} + \nu_2 {\bf 2^+} s_{51} + (d-\nu_{123455}) 
-\left( \nu_1 {\bf 1^+} + \nu_2 {\bf 2^+} + \nu_3 {\bf 3^+} + \nu_4 {\bf 4^+}
 \right) {\bf 5^-} = 0  \\
&& \hspace{-1cm}
\nu_4 {\bf 4^+} s_{45} + \nu_3 {\bf 3^+} s_{12} + (d-\nu_{112345}) 
-\left( \nu_2 {\bf 2^+} + \nu_3 {\bf 3^+} + \nu_4 {\bf 4^+} + \nu_5 {\bf 5^+}
 \right) {\bf 1^-} = 0  \\
&& \hspace{-1cm}
\nu_5 {\bf 5^+} s_{51} + \nu_4 {\bf 4^+} s_{23} + (d-\nu_{122345}) 
-\left( \nu_1 {\bf 1^+} + \nu_3 {\bf 3^+} + \nu_4 {\bf 4^+} + \nu_5 {\bf 5^+}
 \right) {\bf 2^-} = 0 
\end{eqnarray} 
where we denote the invariant masses $s_{ij} \equiv p_{ij}^2$. 
We also note that the topology vanishes if any three adjacent propagators 
are raised to non-positive powers. 

After a preliminary run with no masking algorithms  we find 
11  master integrals:
\begin{itemize}
\item the pentagon integral $V(1, 1, 1, 1, 1),$
\item the box integals $V(1, 1, 1, 1, 0),$ $V(1, 1, 1, 0, 1),$ 
$V(1, 1, 0, 1, 1),$ $V(1, 0, 1, 1, 1),$ $V(0, 1, 1, 1, 1),$ and 
\item the bubble integrals $V(1,0,1,0,0),$ $V(1,0,0,1,0),$ $V(0,1,0,1,0),$
$V(0,1,0,0,1),$ and  $V(0,0,1,0,1).$
\end{itemize}
Remarkably, by using the algorithm for masking reduced expressions, the 
step of Gauss elimination is very fast and can be performed for far more 
complicated integrals than the ones that are practically needed; it only 
takes a few minutes before a sufficiently large number of integrals 
for QCD five-point amplitudes is solved. However, keeping the variable {\tt MAPLEMAXSUB}
in unreachable values (i.e. switching off the special routines for large coefficients, as we did in all previous examples), forces the routine 
{\tt tidy\_int} to carry the load of all substitutions and simplifications. 
The presence of five independent Mandelstam variables and the dimension 
parameter makes the step of nested substitutions very hard for the 
integrals with highest priority. The routine can  get stalled in computers 
with a memory smaller than 1GB.

We set  the variable {\tt MAPLEMAXSUB} to a value of $100$ in 
the {\tt input\_pentagon.map}. The  
reduction routine {\tt Reducer} is not affected by the new setting, however, 
the nested substitution routine {\tt tidy\_list} runs differently; it does not perform simplifications on any expression with more than  $100$ characters. 
Instead, it replaces  these coefficients with indexed aliases, $\{f[i]\}$. 
The routine  does not process  large coefficients,
and finishes very quickly.  The difficult substitutions for the masked 
expressions $f[i]$  are performed with a new routine 
{\tt melt\_all\_f(".")}. This routine is designed to work with the smallest 
possible memory consumption. Finally, the user can display the seed integrals 
explicitly written  in terms of master integrals, the dimension, and the kinematic parameters, using the {\tt show\_full\_int} routine.

It is worth demonstrating some important technical details for  the reduction. 
We perform the reduction in the directory {\tt /home/AIR/Pentagon}, 
where we have placed the input file {\tt input\_pentagon.map},  
and a script for executing the AIR routines 
{\tt script\_pentagon.map}. As usual, in the input file  we have provided 
the IBP equations, conditions for vanishing 
integrals, and the list of master integrals. 
We also set values for the variables, 
\begin{verbatim}
MAXLENGTH:=100:
MAXSIMPLIFY:=1500:
MAPLEMAXSUB:=100:
VERBOSE:=FALSE:
\end{verbatim}

We now perform the reduction of integrals of the pentagon topology with 
${\cal N}_- \leq 4$, and ${\cal N}_+ = 0$. 
\begin{verbatim}
\home\AIR\Pentagon> maple script_pentagon.map
\end{verbatim}
The script is executed in approximately 10 minutes using approximately 40MB of memory. 
At this point it is worth making some observations about the results of the 
reduction. We launch a MAPLE window, 
load the AIR files for the pentagon topology, and read the result for one 
of the reduced integrals: 
\begin{verbatim}
> currentdir(``/home/AIR/Pentagon'');
> read ``input_pentagon.map'';
> read ``/home/AIR/main.map'';
> show_int(V(1, 1, 1, 1, -4), ``.'');
\end{verbatim}
The result now contains masked expressions, e.g. f[67].Their value is stored in the ICED directory tree. We can see 
the value of the masked expressions using the routine:\\
{\bf
show\_f(i, RESULTS\_DIR);
}\\
which displays the variable f[i] for the integer i. We must also provide 
the reduction directory RESULTS\_DIR
For example, the value of f[12] can be retrieved by typing: 
\begin{verbatim}
> show_f(12, ``.'');
\end{verbatim}
Moreover, the command
\begin{verbatim}
> show_full_int(V(1,1,1,1,-4), ``.'');
\end{verbatim}
will return the full expression of the wanted integral.
The commands 
{\tt show\_int}, {\tt show\_full\_int} and {\tt show\_f}  are very convenient to collect the results 
and export them to other platforms for further calculations. 

We proceed, next, with further  reductions of integrals of the pentagon 
topology with ${\cal N}_- \leq 5$, and ${\cal N}_+ = 0$. To perform this 
reduction we have to prepare a run using the same input file and 
requiring additional seeds generated in  the script file, 
\begin{verbatim}
> SEEDGEN("seeds.map",[1,2,3,4,5],[2,5],[0,5],[0,0]);
\end{verbatim}
The script runs now in approximately 25 minutes making use of 65MB of memory. 
The coefficients involved are now larger, but they were computed very 
efficiently. 

We have performed a number of  reductions  for one loop 
hexagon and heptagon topologies and various two-loop topologies.  
For example,  all loop and phase-space integral topologies 
in~\cite{htotal,hrap} are fully reduced in less than 6 hours.
The reduction of the massless double box topology with ${\cal N}_- \leq 4$, 
and ${\cal N}_+ = 0$ can be performed in about three days; the {\tt Reducer} 
routine was running for approximately 7 hours, the {\tt tidy\_list} routine 
went through in one hour and ten minutes, while the {\tt melt\_all\_f} routine worked for a couple of days. The cross-box massless box topology is reduced 
in approximately four days.  In massive two-loop topologies, we have reduced 
integral topologies for the production of heavy quarks. The double-box 
topology with two massive external legs and a massive propagator (all 
carrying the same mass) is reduced using the masking algorithms in about 32 
hours while the nested substitutions are completed in about 20 days 
for ${\cal N}_- \leq 4$, and ${\cal N}_+ = 0$. 

\section{Conclusions}
\label{sec:conclusions}
 
We have presented a MAPLE program for Automated Integral Reductions 
in perturbative calculations. Our program is based on an 
algorithm introduced by Laporta~\cite{laporta}, and 
uses the method of Gauss elimination for solving large systems of  equations. 
The program  can reduce generic  loop or phase-space 
integrals or other functions (like hypergeometric functions) 
which  satisfy coupled algebraic identities. 

The main obstacle in multiloop reductions is the large size 
of the symbolic expressions. We have implemented two algorithms in our program 
which organize the reduction more efficiently and reduce the amount of 
computations.
The routines mask reduced expressions  or large integral coefficients. 
This enables solving the system of IBP equations without performing all 
substitutions explicitly. The computationally intensive task 
of nested substitutions is performed only after the procedure of 
Gauss-elimination is completed, and for only a small fraction of expressions 
which appear in the final results. 

Reduction algorithms cannot be extended to arbitrarily large calculations
due to finite computing resources. We believe, however, that many 
phenomenologically interesting problems are tractable using  AIR.
There are a few improvements that we expect to make in 
future releases, implementing  a more flexible database for storing 
the equations, and including more efficient algorithms for performing 
nested substitutions and simplifying large expressions. The methods described 
in~\cite{weinzierl} are appropriate  for achieving this goal.

\acknowledgments

We are grateful to Frank Petriello for very valuable suggestions, crucial 
observations, and  for his contributions in developing parts of AIR.  
We would like  to thank Lee Garland, Nigel Glover, Thanos Koukoutsakis, 
Carlo Oleari, and Maria Elena Tejeda-Yeomans  for insightful  discussions. 
We would also like to thank Alex Mitov for his detailed feedback and valuable 
suggestions and Thomas Becher, Lance Dixon, Kirill Melnikov and Marc 
Schreiber for their suggestions and encourangament.

\end{document}